\RequirePackage{fix-cm}
\documentclass[smallextended,sort&compress]{svjour3}       
\usepackage{amsmath}
\usepackage{amssymb}
\usepackage{graphicx}
\usepackage{dsfont}
\usepackage{slashed}
\usepackage{cite}

\begin{document}

\title{Electromagnetic baryon form factors in the Poincar\'e-covariant Faddeev approach}

\author{Reinhard Alkofer \and Gernot Eichmann \and H\`elios Sanchis-Alepuz \and Richard Williams}

\authorrunning{R.\ Alkofer, G.\ Eichmann, H.\ Sanchis-Alepuz, and R.\ Williams}

\institute{R. Alkofer  \at Institute of Physics, University of Graz, Universit\"atsplatz 5, A-8010 Graz, Austria\\
		   \email{reinhard.alkofer@uni-graz.at}\\
		   \and
		   G. Eichmann \and H. Sanchis-Alepuz \and R. Williams \at Inst.\ of Theoretical Physics I, Justus-Liebig-Universit\"at Giessen, D-35392 Giessen, Germany \\
}
\date{Received: date / Accepted: date}

\maketitle

\begin{abstract}
Baryons are treated as three-quark systems using QCD degrees of freedom in
Poincar\'e-covariant bound-state equations. The quark self-energy as
well as the interaction between quarks are approximated by a  vector-vector
interaction via a single dressed-gluon exchange (rainbow-ladder truncation),
thereby allowing a unified study of quark, meson and baryon properties. Here we
will focus on the calculation of electromagnetic properties of spin-1/2 and
spin-3/2 ground state baryons.
\end{abstract}

\section{Introduction}
The proton is one of the very few stable hadrons. It has a
complicated extended structure, being a bound state of three valence quarks, a
quark-antiquark sea, and glue. Since the valence quarks constitute just a few
percent of the proton's mass it is evident that the bulk of the proton's
structure and properties are dominated by strong interactions and,
correspondingly, Quantum Chromodynamics (QCD). The goal is then to calculate the
proton's properties.

To do this we observe that the proton, and in general all baryons, appear
as poles in the full six-quark Green function, distinguishable by their quantum
numbers. Since the generation of a pole precludes perturbation theory, we are
left with just a few non-perturbative tools. One possibility is Lattice
QCD, which provides for a numerical simulation of the problem on discrete
Euclidean spacetime.  Another method is that of functional methods, in
particular the Dyson--Schwinger, Bethe--Salpeter and Faddeev approach considered here.
In the following we discuss a selection of results for nucleon and $\Delta$ electromagnetic form factors
as well as $N\to\Delta\gamma$ transition form factors.

\begin{figure}[t]
\begin{center}
\includegraphics[width=0.7\textwidth]{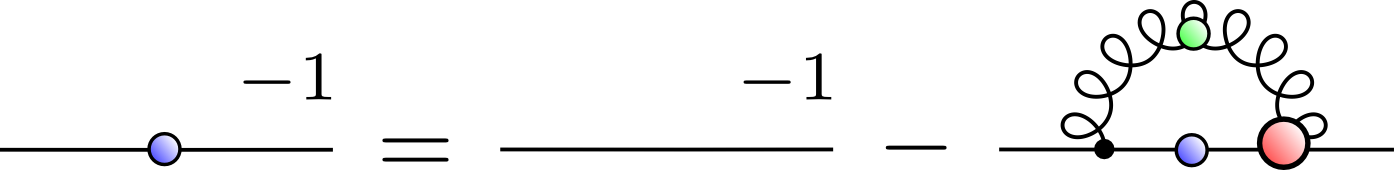}
\caption{The DSE for the quark propagator.}
\end{center}
\end{figure}


%

%

%
%
%
%
%
\section{Relativistic bound-state equations}
The relativistic description of bound states requires as input the propagators
of the constituents, in this case the quarks. The full quark propagator $S$ is
obtained by solving its Dyson--Schwinger equation (DSE)
\begin{align}\label{eqn:quarkDSE}
 S^{-1}(p)=S^{-1}_0(p)+Z_{1f}\int \frac{d^4q}{(2\pi)^4} \gamma^\mu
           D_{\mu\nu}(p-q)\,S(q)\,\Gamma^\nu_{gqq}(p,q)\,,
\end{align}
where $S_0$ is the (renormalized) bare propagator
$S^{-1}_0(p)=Z_2\left(i\slashed{p}+m\right)$, $Z_2$ and $Z_{1f}$
are renormalisation constants, and $m$ is the bare quark mass provided as a
parameter. Its solution requires the full quark-gluon vertex $\Gamma^\nu_{gqq}$
and the gluon propagator $D^{\mu\nu}$.
We make use of the rainbow-ladder (RL) approximation where the
quark-gluon vertex is replaced by its bare form $\Gamma^\nu_{gqq} =\gamma^\nu$.
The corresponding chiral-symmetry preserving quark-quark kernel takes the form
\begin{align}\label{eqn:kernelRL}
K^{\mathrm{2-body}} = \left[\gamma^\mu \otimes \gamma^\nu\right] D_{\mu\nu}(k)\,.
\end{align}

\begin{figure}[!b]
\begin{center}
\includegraphics[width=\textwidth]{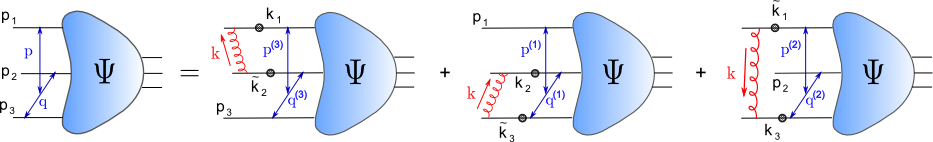}
\caption{The covariant Faddeev equation in rainbow-ladder approximation.}
\end{center}
\end{figure}

Employing a compact matrix notation that implies an integration over continuous variables, the Faddeev equation
is the permuted sum of two-body quark-quark kernels $K^{\mathrm{2-body}}$ and an
irreducible three-body kernel~\cite{Eichmann:2009qa,SanchisAlepuz:2011jn}:
\begin{align}\label{eqn:compactBSE}
\Psi = \big[K^{\mathrm{3-body}}\big] \,G_0\,\Psi + \sum_{a=1}^3
       \big[K^{\mathrm{2-body}}\big]_{(a)}\,G_0\,\Psi\,.
\end{align}
Here, $a$ is an index that labels the spectator quark and $G_0$ is the
disconnected product of three full quark propagators $S_i$. The three-body
kernel does not survive the RL truncation, and so we refer to
$K^{\mathrm{2-body}}$ as $K$.

The conserved current that describes the interaction of a three-quark system with a single photon is given by
\begin{align}\label{eq:current_general}
 J^\mu=&~\bar{\Psi}_f\left(G^{\mu}_0 - G_0K^\mu G_0 \right)\Psi_i\,,
\end{align}
where the incoming and outgoing baryon is described by the covariant Faddeev
amplitudes $\Psi_i$ and $\Psi_f$, respectively.  $G^\mu_0$ is a shorthand
notation for the impulse-approximation diagrams
\begin{align}\label{eq:gauged_G0}
 G_0^\mu=&~\left(S_1\,\Gamma^\mu S_1\right)S_2\,S_3+S_1\left(S_2\,\Gamma^\mu
                 S_2\right)S_3+S_1\,S_2\left(S_3\,\Gamma^\mu S_3\right) ,
\end{align}
with $\Gamma^\mu$ the nonperturbative quark-photon vertex. $K^\mu$ represents the interaction of
the photon with the Faddeev kernel; the only diagrams that survive in RL are those
where the two-body kernel is a spectator:
\begin{align}\label{eq:gauged_K}
 K^\mu=&~ \Gamma^\mu_1\,K_{23} + \Gamma^\mu_2\,K_{31} + \Gamma^\mu_3\,K_{12}\,.
\end{align}
The quark-photon vertex is obtained by solving its vertex BSE~\cite{Maris:1999bh}.
The Faddeev amplitudes including their full Dirac--flavor structure are, in turn,
calculated from the three-body Faddeev equation with the interaction
kernel $K$.

\begin{figure}[p]
\begin{center}
\includegraphics[width=0.9\textwidth]{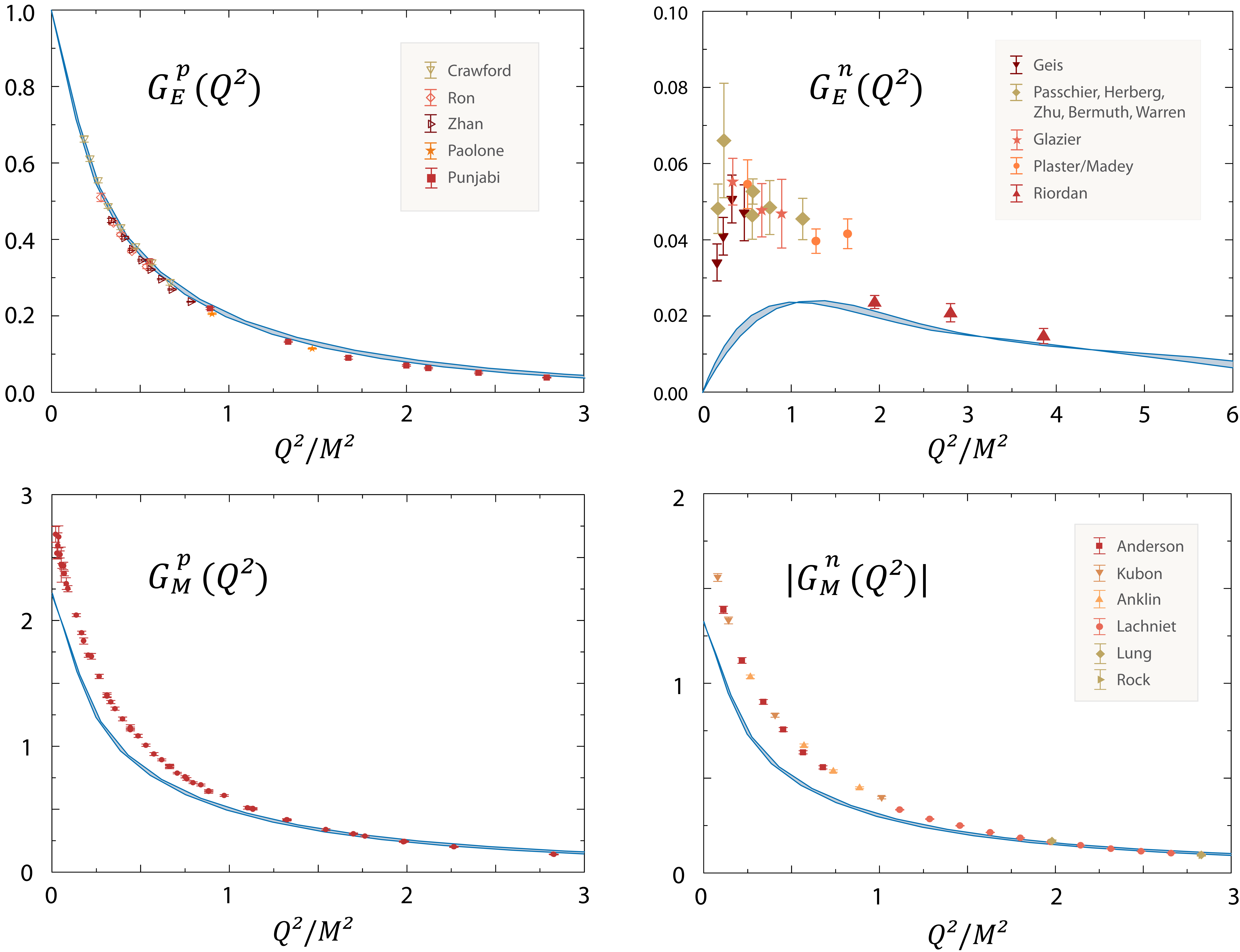}
\caption{Nucleon electromagnetic form factors $G_E$ and $G_M$ for the proton (left panels) and neutron (right panels) compared to experimental data; see~\cite{Eichmann:2011vu} for references.} \label{fig:nucleon-ffs}
\end{center}
\end{figure}
\begin{figure}[p]
\begin{center}
\includegraphics[width=0.9\textwidth]{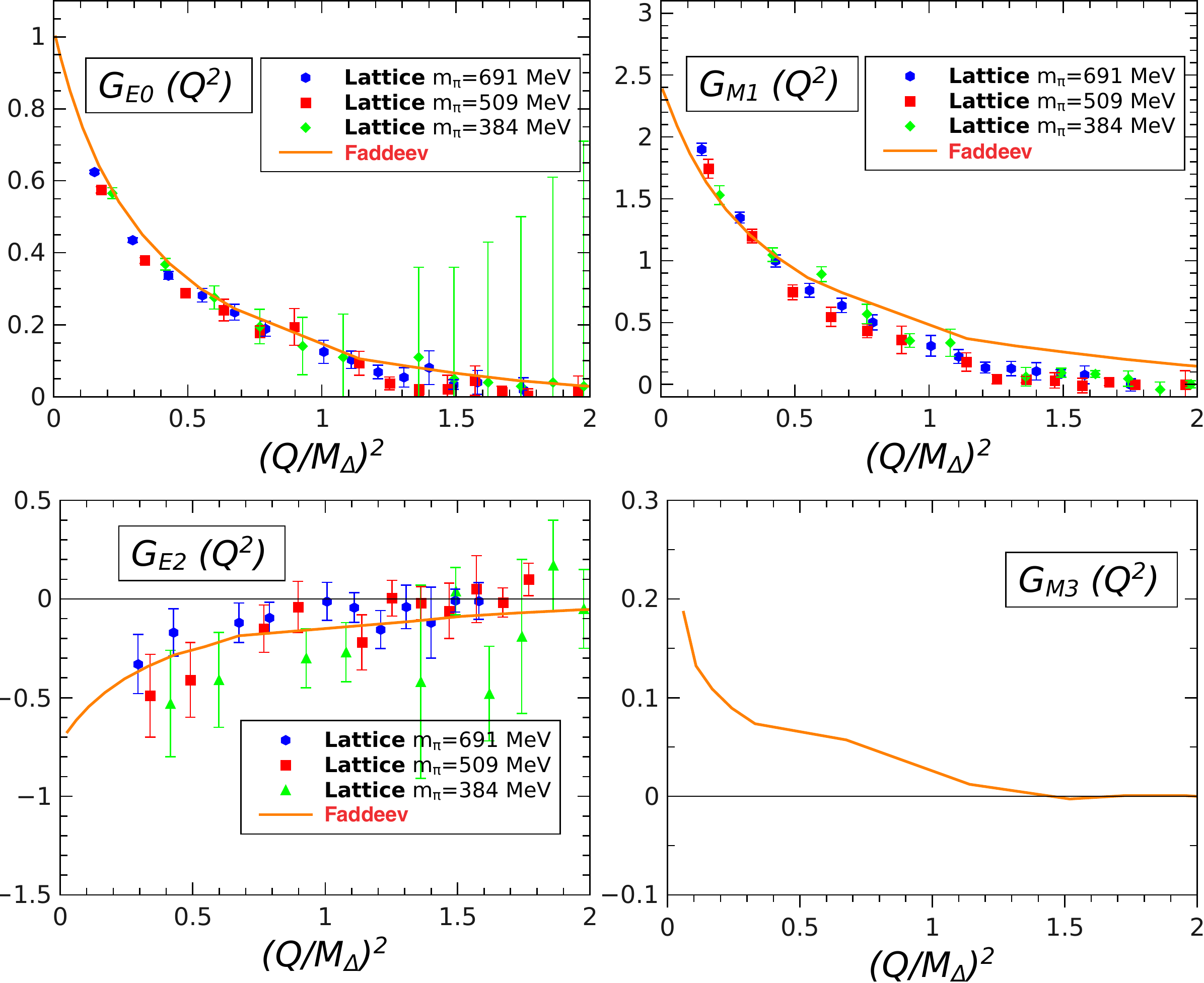}
\caption{$\Delta$ electromagnetic form factors compared to lattice results; see~\cite{Sanchis-Alepuz:2013iia} for references.} \label{fig:delta-ffs}
\end{center}
\end{figure}

\begin{figure}[t]
\begin{center}
\includegraphics[width=\textwidth]{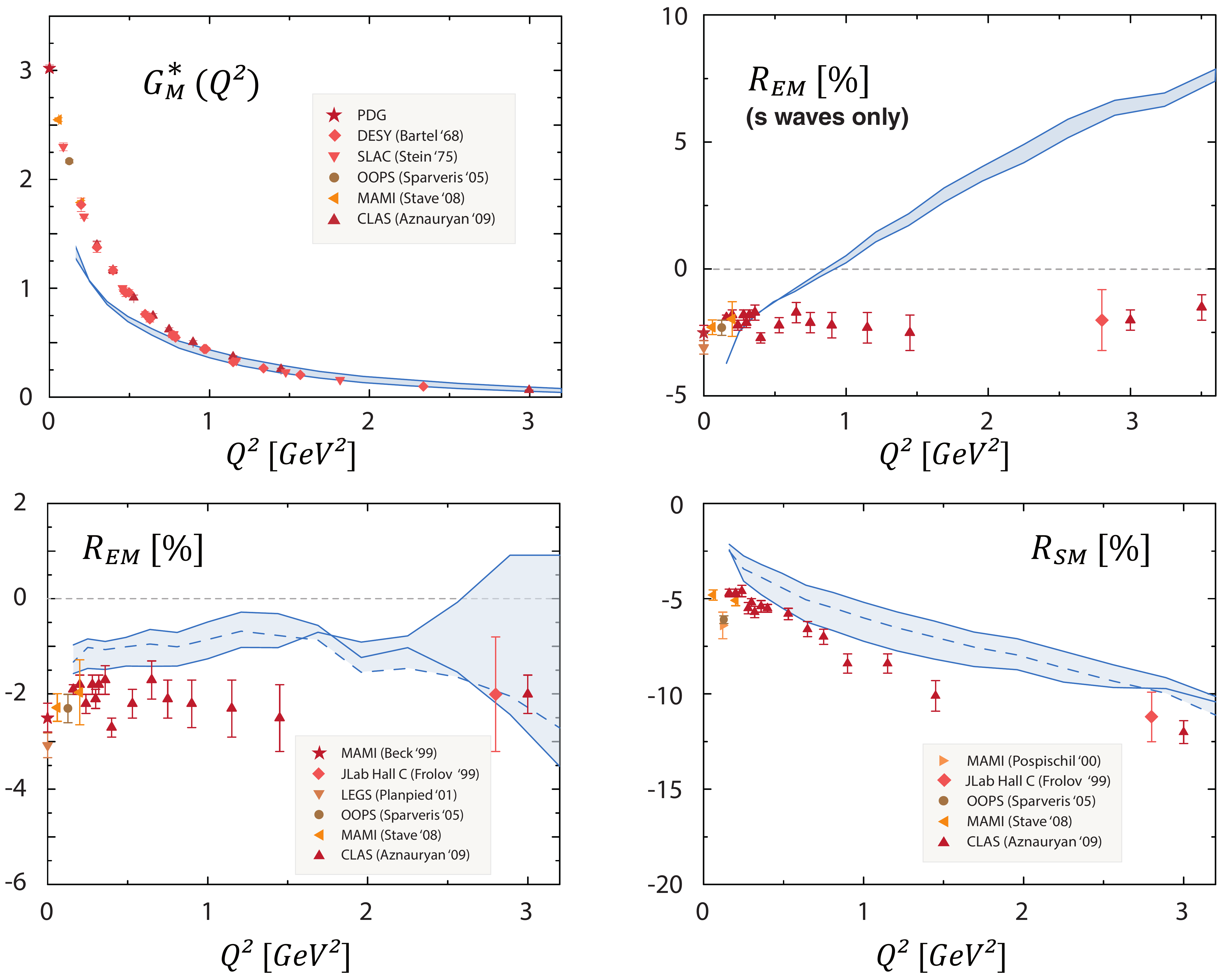}
\caption{$N\to\Delta\gamma$ transition form factors compared to experimental data; see~\cite{Eichmann:2011aa} for references.
         The upper left panel shows $G_M^\star$ and the lower panels the ratios $R_{EM}$ and $R_{SM}$. The upper right panel is the result of an $s-$wave only
         calculation for $R_{EM}$. The dashed lines refer to the central value of the quark-gluon interaction model.} \label{fig:nucleon-delta-ffs}
\end{center}
\end{figure}


%
%
%
%
%
\section{Selected Results}

  The electromagnetic currents that we discuss here are parametrized through dimensionless, Lorentz-invariant form factors (FFs).
  The nucleon's electromagnetic current depends on the two Sachs FFs $G_E$ and $G_M$.
  The $\Delta$ electromagnetic FFs are $G_{E0}$, $G_{M1}$, $G_{E2}$ and $G_{M3}$: electric monopole, magnetic dipole, electric quadrupole and magnetic octupole.
  The $N\to\Delta\gamma$ transition FFs are usually discussed in terms of the magnetic dipole $G_M^\ast$ and the electric and Coulomb quadrupole ratios $R_{EM}$ and $R_{SM}$.

  Our results for the $N$ and $\Delta$ electromagnetic FFs are shown in Figs.~\ref{fig:nucleon-ffs} and~\ref{fig:delta-ffs}.
  Details about the calculations can be found in Refs.~\cite{Eichmann:2011vu,Sanchis-Alepuz:2013iia}; see also~\cite{Eichmann:2011pv} for results on axial FFs.
  In Fig.~\ref{fig:nucleon-delta-ffs} we present first results for the $N\to\Delta\gamma$ transition FFs in the three-body Faddeev approach.
  The comparison with experimental data and lattice results performs rather well in all cases.
  This is quite remarkable given that the only model input in all calculations is the quark-gluon interaction in Eq.~\eqref{eqn:kernelRL},
  with the model dependence indicated by the bands in Figs.~\ref{fig:nucleon-ffs} and~\ref{fig:nucleon-delta-ffs}.
  To some extent this can be attributed to the global symmetries of QCD, including its pattern of spontaneous chiral symmetry breaking, which are respected in every step of these calculations.
  For example, charge conservation at $Q^2=0$ is not imposed by hand but rather a consequence of the underlying Ward-Takahashi identities.

  On the other hand, the absence of structure is visible at low $Q^2$, in particular for the magnetic form factors.
  These are symptoms of missing meson-cloud effects, which would enhance magnetic moments and charge radii close to the chiral limit
  and produce cusp effects in $\Delta$ form factors due to the $N\to\Delta\pi$ decay.
  A RL truncation does not support this; it produces stable bound states that are stripped from their pion cloud and do not decay.
  The comparison in Fig.~\ref{fig:delta-ffs} is illuminating in this regard: the $\Delta$ calculated in lattice QCD is a bound state at unphysical pion masses, below $N\pi$ threshold.
  When plotted as a function of $Q^2/M_\Delta^2$, the lattice results for different pion masses fall into relatively narrow bands that agree well with the Faddeev calculation.

  The importance of pion-cloud effects has also been stressed in the context of the $N\to\Delta\gamma$ transition, because the ratio $R_{EM}$ is sensitive to orbital angular momentum
  in the $N$ and $\Delta$ wave functions~\cite{Pascalutsa:2006up}. The covariant Faddeev amplitude $\Psi$ of the $\Delta$ obtained from Eq.~\eqref{eqn:compactBSE} has a rich structure in terms of 128 Dirac tensors.
  Only four of them are $s$ waves in the $\Delta$ rest frame and resemble the `orbital ground-state' wave functions in the quark model. The remaining ones are relativistic $p$ waves and provide
  orbital angular momentum, together with further $d$ and $f$ waves whose effect is much smaller~\cite{SanchisAlepuz:2011jn}. Similarly, the nucleon's Faddeev amplitude has 64 Dirac structures that can be arranged
  in $s$, $p$ and $d$ waves~\cite{Eichmann:2009qa}. These higher tensor structures in the $\Delta$ are presumably also the culprit for the noisy behavior in $R_{EM}$ that is seen in Fig.~\ref{fig:nucleon-delta-ffs}.

  In any case, if only the $s-$wave components are retained (upper right panel in Fig.~\ref{fig:nucleon-delta-ffs}), 
  the ratio $R_{EM}$ rises sharply towards the asymptotic prediction $R_{EM}\rightarrow +100\%$.
  That this behavior is not seen in the experimental data (which remain negative up to the highest data point at 6 GeV$^2$~\cite{Aznauryan:2009mx})
  is perhaps not too surprising since analogous arguments predict $\mu_p G_E/G_M \rightarrow 1$ for the proton,
  whereas experiments show a clear falloff with a possible zero crossing~\cite{Puckett:2011xg}.
  The $p-$wave admixture in a realistic Faddeev amplitude generated by the RL truncation produces a similar falloff in $G_E/G_M$~\cite{Eichmann:2011vu},
  together with the negative value for $R_{EM}$ in Fig.~\ref{fig:nucleon-delta-ffs}.

  Finally, we note that similar results for these FFs have been obtained in the quark-diquark approach of Refs.~\cite{Eichmann:2009zx,Nicmorus:2010sd,Eichmann:2011aa},
  and to some degree also in simpler quark-diquark models~\cite{Cloet:2008re,Segovia:2013rca}. Although the three-body Faddeev equation in Eq.~\eqref{eqn:compactBSE}
  carries no trace of diquarks by itself, this adds support to the quark-diquark interpretation
  for these lowest-lying baryons. 
  It will be interesting to see whether the picture changes appreciably beyond the rainbow-ladder approximation~\cite{Fischer:2009jm,Williams:2014iea,SanchisAlepuz:2015a}.


%
%
%
%
%

%
%
%
%
%
\section*{Acknowledgements}
R.A.\ is grateful to the organizers of the
{\it International Conference on Exotic Atoms and Related Topics - EXA2014}
for all their efforts which made this conference possible, and for financial support.
  This work has been supported by an
Erwin Schr\"odinger fellowship J3392-N20 from the Austrian Science Fund FWF, by the Helmholtz
International Center for FAIR within the LOEWE program of the State of Hesse, and by the
DFG collaborative research center TR 16. Further support by the  European Union (Hadron
Physics 3 project ``Exciting Physics of Strong Interactions'') is acknowledged.


\begin{thebibliography}{99}


\bibitem{Eichmann:2009qa}
  G.~Eichmann {\it et al.},
  Phys.\ Rev.\ Lett.\  {\bf 104}, 201601 (2010).


\bibitem{SanchisAlepuz:2011jn}
  H.~Sanchis-Alepuz {\it et al.},
  Phys.\ Rev.\ D {\bf 84}, 096003 (2011).


\bibitem{Maris:1999bh}
  P.~Maris and P.~C.~Tandy,
  Phys.\ Rev.\ C {\bf 61}, 045202 (2000).


\bibitem{Eichmann:2011vu}
  G.~Eichmann,
  Phys.\ Rev.\ D {\bf 84}, 014014 (2011).


\bibitem{Sanchis-Alepuz:2013iia}
  H.~Sanchis-Alepuz, R.~Williams and R.~Alkofer,
  Phys.\ Rev.\ D {\bf 87}, 096015 (2013).


\bibitem{Eichmann:2011aa}
  G.~Eichmann and D.~Nicmorus,
  Phys.\ Rev.\ D {\bf 85}, 093004 (2012).


\bibitem{Eichmann:2011pv}
  G.~Eichmann and C.~S.~Fischer,
  Eur.\ Phys.\ J.\ A {\bf 48}, 9 (2012).


\bibitem{Pascalutsa:2006up}
  V.~Pascalutsa, M.~Vanderhaeghen and S.~N.~Yang,
  Phys.\ Rept.\  {\bf 437}, 125 (2007).

\bibitem{Aznauryan:2009mx}
  I.~G.~Aznauryan {\it et al.}  [CLAS Collaboration],
  Phys.\ Rev.\ C {\bf 80}, 055203 (2009).

\bibitem{Puckett:2011xg}
  A.~J.~R.~Puckett {\it et al.},
  Phys.\ Rev.\ C {\bf 85}, 045203 (2012).


\bibitem{Eichmann:2009zx}
  G.~Eichmann, PhD Thesis, U. Graz,
  arXiv:0909.0703 [hep-ph].


\bibitem{Nicmorus:2010sd}
  D.~Nicmorus, G.~Eichmann and R.~Alkofer,
  Phys.\ Rev.\ D {\bf 82}, 114017 (2010).


\bibitem{Cloet:2008re}
  I.~C.~Cloet {\it et al.},
  Few Body Syst.\  {\bf 46}, 1 (2009).


\bibitem{Segovia:2013rca}
  J.~Segovia, C.~Chen, C.~D.~Roberts and S.~Wan,
  Phys.\ Rev.\ C {\bf 88}, 032201 (2013).


\bibitem{Fischer:2009jm}
  C.~S.~Fischer and R.~Williams,
  Phys.\ Rev.\ Lett.\  {\bf 103}, 122001 (2009).


\bibitem{Williams:2014iea}
  R.~Williams,
  arXiv:1404.2545 [hep-ph].

\bibitem{SanchisAlepuz:2015a}
  H.~Sanchis-Alepuz, R.~Williams, \emph{in preparation}.

\end{thebibliography}
\end{document}